\documentstyle[prb,aps,epsf]{revtex}
\begin{document}
\draft

\twocolumn[\hsize\textwidth\columnwidth\hsize\csname@twocolumnfalse%
\endcsname

\title{RKKY interaction in the nearly-nested Fermi liquid}

\author{D.N. Aristov $^{1,2}$, S.V. Maleyev $^2$}

\address{ 
 $^1$ Laboratoire L{\'e}on Brillouin, CE-Saclay, 
91191 Gif-sur-Yvette Cedex, France. } 
\address{$^2$ Petersburg Nuclear Physics Institute,
Gatchina, St. Petersburg 188350, Russia}
\date{\today}
\maketitle

\begin{abstract}  
We present the results of analytical evaluation of the indirect RKKY 
interaction in a layered metal with nearly nested (almost squared) 
Fermi surface. The final expressions are obtained in closed form 
as a combination of Bessel functions. We discuss the notion of the 
 ``$2k_F$'' oscillations and show that they occur as the far 
asymptote of our expressions. We show the existence of 
the intermediate asymptote of the interaction which 
is of the sign-reversal antiferromagnetic type and is the only term
surviving in the limit of exact nesting.  
A good accordance of our analytical formulas with numerical findings  
is demonstrated until the interatomic distances. 
The obtained expressions for the Green's functions extend the previous 
analytical results into the region of intermediate distances as well.
\end{abstract}

%%%%%%%%%%%%%%%%%%%%%%%%%%%%%%%%%%%%%%%%%%%%%%%%%%%%%%%%%%%%
]

%\narrowtext
  
\def\bk{{\bf k}}
\def\bq{{\bf q}}
\def\br{{\bf r}}

\def\mass{\tensor{ {\rm m}}}
\def\ve{\varepsilon}

The Ruderman-Kittel-Kasuya-Yosida (RKKY) interaction was found to play
an important role in various problems involving the interaction of
localized moments in a metal via polarization of conduction electrons.
The spatial dependence of this interaction 
for the spherical Fermi surface (FS) in three dimensions was
obtained in Ref.\ \cite{rkky} about 40  years ago. It was demostrated there
that at large distances $r$ 
the interaction decays as $r^{-3}$ and has the $2k_F$ oscillations with
the Fermi momentum $k_F$. Later Roth, Zeiger and Kaplan \cite{rkky-ani}
have generalized this 
result for the case of non-spherical FS. The interaction was
represented as a series in $1/r$ and the existence of the direction-dependent
period of oscillations $(2k_F^\ast)^{-1}$was shown.    
 
The limitation of this latter result is the inapplicability of the theory
at small distances $k_F^\ast r\lesssim 1$, i.e.\ in the region where the RKKY 
interaction is mostly significant. 
It could be among the plausible reasons 
for the nowaday situation, when the notion of $2k_F$ oscillations is widely
explored while the actual meaning of this term for the non-spherical 
FS remains unclear in many cases.   
Numerical calculations are instrumental \cite{FeCr}
 to extend our understanding of
non-spherical FS ever since 1957, but a theoretical understanding even 
on a qualitative level is very important. 

In a recent paper one of the authors
presented the method which enabled  to obtain in 
a simple manner the RKKY interaction for the spherical FS and any 
dimensionality of the system. The closed expressions were found both in 
$r-$ and $q-$ representations. \cite{rkky-anyD}  

In the present paper we generalize this  
method to arrive to  the closed
analytical expressions for the RKKY interaction in a layered (2D) metal with 
highly non-spherical (nearly nested) Fermi surface. Our main results 
could be summarized as follows. 

\noindent
i) The RKKY interaction for this type of the FS consists of 
three parts. The first one stems from the flat regions in the fermionic 
spectrum. This term decays mostly as $1/r$ and is important only along 
the diagonals  $ x = \pm y$ in the 
$\br-$space.  

\noindent
ii) The second part of 
  the RKKY interaction is contributed by the 
vicinities of the points $(0,\pi)$ and $(\pi,0)$. These are the 
saddle points in the fermionic dispersion giving rise to the 
van Hove singularities.     
  At $r\to \infty$ 
this  part of RKKY  takes its exact correspondence 
with the previous findings. \cite{rkky-ani} It turns out however that it 
is the far asymptote of the interaction. 

\noindent
iii) At $k_F^\ast r \lesssim 1$ 
the third part of the interaction comes into play. 
This intermediate asymptote arises as an interference 
between contributions from the vicinities of the different van Hove points. 
This term possesses the overall prefactor 
$\cos ({\bf Q} \br)$ with the spanning wave-vector ${\bf Q} = (\pi,\pi)$.
 In the nearly nested situation   this last term is 
present down to the interatomic distances and strongly favors the 
commensurate antiferromagnetic ordering of the localized moments.

\noindent
iv) The last but not least. 
We show both analytically and numerically that our expressions are valid
near the interatomic distances. This result could be 
 qualitatively  explained in the following way. 
It is clear that to evaluate the interaction at the  
distances $r$ one should know the details of the 
fermionic dispersion on a scale $1/r$ in $\bk-$space.
 Hence
the finest details of the Fermi surface are of importance at the largest 
distances. 
One may also  
conclude that even rough knowledge of the spectrum 
over the whole Brillouin zone is enough for a good description of the 
corresponding quantities at the interatomic distances in $\br-$space.    
This is exactly what our method does 
by grasping the key features of the FS.

The rest of the paper is organized as follows. 
We formulate the problem and introduce the basic ingredients of our 
treatment in Sections  I and II. The Green's functions resulted from the 
van Hove points of the spectrum are found in Section III. Their contribution 
to the RKKY interaction is analyzed in Section IV. The role of the flat 
parts of the spectrum is discussed in Section V. We make the concluding 
remarks in the Section VI of the paper.

 \section{general formalism}

We begin with conventional form of the exchange interaction
between the localized moment ${\bf J}$ and electron spin density
${\bf s}(\br)$ :

     \begin{equation}
     V(\br)=
     -A {\bf J}({\bf R}) {\bf s}(\br) \delta({\bf R}-\br)
     \end{equation}
Here $A$ is the exchange coupling constant.  The RKKY
interaction between two localized moments via the conduction
electrons may then be written in the following form

     \begin{equation}
     \label{rkky-def}
     H_{RKKY}= - \frac12 A^2 {\bf J}_1{\bf J}_2 \chi(
     \br_{1},\br_{2})\ .
     \end{equation}
where $r-$dependent part of the interaction
coincides with the Fourier transform of the non-uniform static
susceptibility $\chi(q)$ (Lindhard function) and is usually
written in the form :

     \begin{eqnarray}
     \chi(\br_{1},\br_{2}) &=&
     \frac{v_0^2}{(2\pi)^6}
     \int d^3\bk\, d^3{\bf q}\,
     \frac {n_k- n_q} {\varepsilon_q -\varepsilon_k}
     e^{i({\bf q} - \bk)(\br_{1}-\br_{2})}
     \label{rkk-conv}
%     \\   &&   \times
%     u_k^\ast(\br_1) u_k (\br_2)
%     u_q(\br_1) u_q^\ast (\br_2)
%     \nonumber
     \end{eqnarray}
with
the unit cell volume $v_0$ and the Fermi function $n_k
=(\exp(\varepsilon_k/T) + 1)^{-1}$.  For our purpose, it is more
convenient however to represent the above expression in the
equivalent form

	\begin{equation}
     \chi(\br_{1},\br_{2}) =
     - T \sum_n
     G(i\omega_n, \br_{1},\br_{2} )
     G(i\omega_n, \br_{2},\br_{1} )
	\label{rkk-inter}
 	\end{equation}
here Matsubara frequency $ \omega_n = \pi T(2n+1)$ and $G$ is the 
electronic Green's function.   In the case
of low temperatures considered below we use the limiting
relation $T \sum_n \to \int_{-\infty}^\infty d\omega/(2\pi)$.

The electronic Green's function is given by

     \begin{equation}
     G(i\omega, \br_{1},\br_{2}) =
     \frac{v_0}{(2\pi)^3}
     \int d^3\bk
     \frac{\exp(i\bk(\br_{1}-\br_{2}))}
     {i\omega -\varepsilon_k}
     u_k(\br_1) u_k^\ast (\br_2)
     \label{g-def}
     \end{equation}
with the periodic Bloch function $u_k(\br)$. 
Generally speaking the $u-$functions should be inserted into 
(\ref{rkk-conv}), and it is the case of free electrons only when 
the RKKY interaction depends on
the absolute value of the distance $\br_{1}-\br_{2}$.  We consider
below the tight-binding form of the electronic Hamiltonian which
implies the following form of $u_k(\br)$ \cite{aaa}:

     \begin{equation}
     u_k(\br) =
     \sum_n e^{i\bk ({\bf a}_n -\br) }
     \varphi(\br- {\bf a}_n) ,
     \label{bloch}
     \end{equation}
here the Wannier function $\varphi(\br)$ rapidly decays
with distance and is close to the atomic wave-function at
small $\br$. In view of this rapid decrease on the scale of
interatomic distances, it is possible to neglect the dependence
of  $u_k(\br)$ on $\bk$ within the first Brillouin
zone for most positions of $\br$ in the unit cell. 
Therefore we may replace $u_k(\br)$ by $u_0(\br)$
and assume the exchange coupling is appropriately redefined, 
$A \to A|u_0(\br)|^2$. It is clear that upon this
redefinition the Green's function (\ref{g-def}) depends only on
the difference $\br \equiv \br_{1} - \br_{2}$.

In view of forthcoming consideration we stress that the above replacement 
of the amplitude of Bloch function
is the only uncontrolled appoximation of our treatment. 
The relevant calculations \cite{Bel-Ch}
show that  $\varphi(\br)$ 
for the nearest neighbors is of order of magnitude smaller than $\varphi(0)$. 
Therefore we expect that omitting the dependence
of  $u_k(\br)$ on $\bk$ might cause only minor corrections 
to our results.

\section{nearly  nested Fermi surface}
 
We consider a two-dimensional case of almost nested
Fermi surface, with the quasiparticle dispersion given by

	\begin{equation}
     \ve_{\bk} =
     -2t (\cos  k_x + \cos  k_y ) - \mu,\quad |\mu|  \ll t 
     \label{disp-nest}
	\end{equation}
with $t> 0 $. 
This tight-binding form of the spectrum particularly appeared in
different models related to the high-$T_c$ phenomenon.
\cite{htsc} Henceforth we let the lattice parameter to be unity.

We will regard the different parts of the spectrum
(\ref{disp-nest}) on the different manner.
We  divide the whole FS onto four parts,
which are as follows.

{\em The vicinities of the saddle points of the
spectrum.} The dispersion near the points $(0,\pm\pi)$ and
$(\pm\pi,0)$ are  given by

	\begin{equation}
     \ve_{ (0,\pm\pi) +\bk} \simeq
     t (k_x^2 - k_y^2 ) - \mu + O(t k^4) ,
      \label{exp1}
	\end{equation}
     \begin{equation}
     \ve_{(\pm\pi,0) + \bk } \simeq
     -t (k_x^2 - k_y^2 ) - \mu + O(t k^4) ,
      \label{exp2}
	\end{equation}
respectively.
 
{\em The flat parts of the Fermi surface.}
In the vicinity of the pair of wave-vectors $\pm(\pi/2,\pi/2)$
the dispersion takes the form similar to the one-dimensional case near 
half-filling \cite{Luther}:
 
     \begin{equation}
     \ve_{\bk \pm(\pi/2,\pi/2)} \simeq
     \pm 2t (k_x + k_y ) - \mu + O(t k^3) .
      \label{exp3}
	\end{equation}
The pair $\pm(\pi/2,-\pi/2)$
is characterized by the similar dispersion law  
	\begin{equation}
     \ve_{\bk \pm(\pi/2,-\pi/2) } \simeq
     \pm 2t (k_x - k_y ) - \mu + O(t k^3) .
     \label{exp4}
	\end{equation}

The   expansions (\ref{exp1})-- (\ref{exp4}) are valid until 
$k\lesssim 1$. In
its turn, it means that dropping the higher terms of the
  expansions, one hopes to obtain the correct form of RKKY
interaction at $r\gtrsim 1$. Below we revise this statement
and  compare  our analytic results with the
numerical findings.

We see that the whole vicinity of the FS can
naturally be divided into parts of the mainly two-dimensional 
hyperbolical and
 linear character of dispersion.  Therefore we can
represent the Green's function as a sum of eight different
contributions which is symbolically written in the form

     \begin{eqnarray}
     G(i\omega, \br  ) &=& 
     \sum_{\bk_0}
     G_{\bk_0 }(i\omega, \br )
     \label{G-total}
     \end{eqnarray}
where $\bk_0 = \pm(0,\pi), \pm(\pi,0), (\pm\pi/2, \pm\pi/2) $ and
{\em the partial Green's function} $ G_{\bk_0 }(i\omega,
\br ) $ originates from the integration over the part of the
Brillouin zone with particular form of dispersion law :

     \begin{equation}
     G_{\bk_0 } (i\omega, \br )  =
     \frac{v_0}{(2\pi)^2}
     \int_{k\lesssim 1} d^2\bk
     \frac{\exp(i(\bk+\bk_0)\br)}
     {i\omega -\varepsilon_{\bk+\bk_0}} .
     \label{gpart-def}
     \end{equation}  
It should be noted that this decomposition is a general one and may be 
done in all cases when the FS possesses the saddle points and flat parts. 
Hence our consideration of the nearly nested FS may be regarded as a 
 particular example of the  analysis of the RKKY interaction for 
the strongly non-spherical FS.

\section{ the Green's function from the saddle points in 
the spectrum}

In this Section we consider the contribution to the Green's function from the
vicinities of the saddle points of the spectrum only. The flat parts 
are analyzed in Section \ref{app:flat}. As we will show below the saddle 
points determine the RKKY interaction almost at all directions in $r-$space 
except the regions near the diagonals $x=\pm y$ where the contribution from 
the flat parts of the FS becomes important.
 
At the first step we use the following auxiliary representation of the
Green's function  :

     \begin{equation}
     \label{g-tau1}
     G(i\omega, \br ) =
     \frac{ e^{-i\alpha} }{(2\pi)^2}     \!
      \int\limits_0^{\infty} \! d\tau \!
     \int \!\!\!
     d^2\bk
     \exp \!\left[
     i\bk\br + \tau e^{i\alpha}
     \left[i\omega - \varepsilon_k \right]
     \right]
	\end{equation}
where we put $\alpha = sign(\omega) \pi/2$. 
It is evident that  $e^{i\alpha} = i\, sign(\omega)$ but this formal trick
facilitates the following analysis of the complex-valued expressions. 

The evaluation of the saddle point contribution to the Green's funciton can be 
made in more general form applicable for the parts of the electronic spectrum 
near the so-called stationary points \cite{aaa}. 
By definition these are the points $\bk_0$ where
$\varepsilon_{\bk}$ is well approximated by the expression

     \begin{equation}
     \label{stpoint}
     \ve_{\bk_0+ \bk}
     \simeq  \frac12 \bk\, \mass^{-1} \, \bk
      - \mu ,
     \end{equation}
with the generally anisotropic tensor of masses $\mass$ and some
effective chemical potential $\mu$. We assume the applicability
of (\ref{stpoint}) for the wave-vectors $\bk$ not exceeding
some scale $\kappa$, comparable to inverse lattice parameter. In
other words the significant parts of the FS may be approximately mapped
by (\ref{stpoint}).

 In view of (\ref{stpoint}) the integration over $\bk$ 
in (\ref{g-tau1}) becomes 
Gaussian one and is easily performed. The only complication here 
comparing to the case of free electron gas \cite{rkky-anyD} is
the finite value of $\kappa$ which restricts the validity of the subsequent 
equations by $r \gtrsim \kappa^{-1}$;  we discuss it in more detail below.    

 Thus we obtain the following expression for the partial
contribution to Green's function from the vicinity of $\bk_0$ 

     \begin{eqnarray}
     G_{\bk0}(i\omega, \br ) &=&
     - e^{-i\alpha}\frac{ \sqrt{|det\, \mass}|}{2\pi} e^{i\bk_0\br}
     \nonumber \\
     &\times&
     \int_0^{\infty} \frac{d\tau}{\tau}
     \exp[ \tau z e^{i\alpha} -
     \frac{\rho}{2\tau} e^{-i\alpha}  ]
     \label{g-tau2}
     \end{eqnarray}
where  $z = \mu + i\omega$
and $\rho = \br\, \mass \,\br$ is the square of distance {\em
in the metrics defined by the mass tensor}.   
The last integral  is expressed via the modified Bessel
(Macdonald) function, \cite{GR} namely

     \begin{equation}
     G_{\bk0}(i\omega, \br ) =
     - e^{-i\alpha}\frac{\sqrt{|det\, \mass|}}{\pi} e^{i\bk_0\br}
     K_0 (\sqrt{-2z\rho})
     \label{g-mcdo}
     \end{equation}

In the equation (\ref{g-mcdo}) the branch of $\sqrt{-2z\rho}$ is
chosen from the condition of its positive real part.  In
particular case of $2\mu\rho >0$, this latter condition means
that the argument of Macdonald function $K_0 (\sqrt{-2z\rho})
$ has a discontinuity at $\omega = 0$, \cite{GR}

     \begin{equation}
     K_0 (\sqrt{-2z\rho}) =
     \left\{
     \begin{array}{rl}
     \frac{\pi i}2  H_0^{(1)}(\sqrt{2\mu\rho}), &
     \quad \omega/\mu \to +0 \\
     -\frac{\pi i}2  H_0^{(2)}(\sqrt{2\mu\rho}), &
     \quad \omega/\mu \to -0
     \end{array}
     \right.
     \label{K2H}
     \end{equation}
where $H_0^{(1,2)}(x)$ are Hankel functions. Note that in the
case of spherical Fermi surface the condition $2\mu\rho \equiv
k_F^2 r^2 > 0 $ is fulfilled automatically. For the 
electronic type of dispersion  $\mass >0$ and $\mu >0$ while  
for the hole-like dispersion law  one has $\mass<0 $ and $\mu< 0$.  
It is clear from Eq.(\ref{K2H}) that for non-spherical Fermi surface 
the effective Fermi momentum is given by $\sqrt{2\mu\rho }= k_F^\ast r$
and may strongly depend on the direction in real space 
(cf.\ Eq.(\ref{def-kF}) below).

Let us discuss now  the region of 
applicability of the  expression (\ref{g-mcdo}).
The Gaussian integration in (\ref{g-tau1}) 
for $\ve_\bk$ given by (\ref{stpoint}) is
justified upon two conditions. First, the center of
quadratic form in ${\bf k}$ which is $\tau^{-1} \br \,\mass$
should lie within the circle of radius $\kappa$, where
the expansion (\ref{stpoint}) is applicable.
Second, one should demand the criterion $\tau |{\bf
\kappa}\,\mass^{-1} \,{\bf \kappa} |\gg 1$ to be
satisfied , to ensure 
the Gaussian value (\ref{g-tau2}).
In the principal axes $j$ of $\mass$ these
criteria can be combined as follows : 
       \begin{equation}
       \tau \gg \max\left[ \frac{|r_jm_j|}{\kappa},
       \frac{|m_j|}{\kappa^2}\right]
       \label{star}
       \end{equation}  
At this point our analysis is somewhat different for the cases 
$ k_F^\ast r = \sqrt{2\mu\rho}\gg 1 $ and $ k_F^\ast r \lesssim 1 $. 

In the first case of largest $r$ the principal contribution to the integral 
(\ref{g-tau2}) is delivered by  $\tau = \tau_0 (1 + O(1/\sqrt{2z\rho}) )$ with
$\tau_0 = \sqrt{\rho/2z} = k_F^\ast r/(2\mu)$ \cite{fnote2}. 
Simple arguments show then that for the ellipsoidal FS the criterion 
({\ref{star}) is always fulfilled, coinciding with the obvious demand 
$k_F^\ast \ll \kappa$. For the spectrum with the saddle point 
the condition ({\ref{star}) 
may be violated at large distances and near the nodes $\rho\simeq 0$. 
Let the angle $\varphi$ 
be measured from the $\hat x$-axis in $\br$-plane.  
Then $ k_F^\ast $ is given by (\ref{def-kF})
and we lose the applicability of (\ref{g-mcdo}) at 
       \begin{equation}
       \frac{\sqrt{|t/\mu|} }{r} \ll \sqrt{|\varphi \pm \pi/4|} 
         \lesssim     \frac{\sqrt{|\mu/t|} }{a},
       \label{star2}
       \end{equation} 
i.e.\ within a narrow sector along the diagonals if 
$r/a \gg |t/\mu| \gg 1$.      

In the case of $ k_F^\ast r \lesssim 1 $ it is the region 
$|\rho|\lesssim \tau \lesssim |z|$ which is essential in (\ref{g-tau2}). 
Substituting  the lower boundary  $\tau = |\rho| $ into  (\ref{star}) we 
come to the evident condition $\kappa r > 1 $ for the ellipsoidal FS. 
For the FS with the saddle point the criterion (\ref{star}) 
is again violated near the node of $\rho$. For our particular form of the 
spectrum the Eq.(\ref{g-mcdo}) is applicable outside the band along the 
diagonals $x=\pm y$ : 

       \begin{equation}
        r |\varphi \pm \pi/4| > \kappa^{-1} \sim a.
       \label{star3}
       \end{equation}
We depict the above regions of applicability on the Figure \ref{fig:regions}. 

Let us now consider the particular form of the Green's functions arising in the
case of the tight-binding spectrum (\ref{disp-nest}).  

The integration over $\bk$ being restricted to the first Brillouin 
zone leads to the additional phase factor in (\ref{g-mcdo}) as discussed 
in Appendix \ref{app:bzb}. The desired expressions acquire the form :

     \begin{eqnarray}
     G_{(0,\pi)}(i\omega, \br ) &=&
     - \frac{ i sign(\omega) }{2\pi t}
     K_0 (\sqrt{-2z\rho})  e^{i \pi |y| sign(\omega ) }
     \label{gf0pi}  \\
     G_{(\pi,0)}(i\omega, \br ) &=&
     - \frac{ i sign(\omega)}{2\pi t} 
     K_0 (\sqrt{2z\rho}) e^{i \pi |x| sign(\omega)}
     \label{gfpi0}     
     \end{eqnarray}
where $\rho = ( x^2 - y^2 )/ 2t $ and  $\sqrt{|det\,\mass|}$ is 
replaced by its actual value $1/2t$.

\section{evaluation of the RKKY interaction}
\subsection{integer values of r}

We note that if the values of coordinates $x,y$ 
in (\ref{gf0pi}), (\ref{gfpi0}) coincide with the integer 
numbers of lattice periods then one has  $ e^{i \pi |x| sign(\omega )} = 
 e^{i \pi x }$, $ e^{i \pi |y| sign(\omega )} = 
 e^{i \pi y }$ and $ e^{2i \pi x } =  e^{2i \pi y } = 1$.  
We consider this simpler case first, 
while the case of non-integer $x,y$ is discussed 
in the next subsection.

 Away from the diagonals $x=\pm y$ in $\br$-space, we have 
from (\ref{rkk-inter}) the following expression : 

     \begin{eqnarray}
     \chi(\br) &=&
       \frac {1}{ 4\pi^2 t^2}
     \int_{\mu- i\infty}^{\mu+i\infty} \frac{dz}{2\pi i}
     \left[ K_0^2 (\sqrt{-2z\rho}) + K_0^2 (\sqrt{2z\rho})
     \right. \nonumber \\ && \left.
     + 2  e^{i\pi(x+y)} K_0(\sqrt{-2z\rho})  K_0(\sqrt{2z\rho}) 
     \right] 
     \label{rkk-ne-0}
     \end{eqnarray}   
Here we redefined the variable of integration $\omega \to z = \mu+i \omega$. 

As we discussed above if $\mu\rho > 0$ then the function $K_0(\sqrt{-2z\rho})$
has the discontinuity (\ref{K2H}) at $z =\mu$.
On the contrary
 the second term in (\ref{rkk-ne-0}) is continuous function 
of $z$ in this case and the corresponding integral is zero, because one can
shift the integration contour to $z\to +\infty$ wherein
 $K_0(z) \propto e^{-z}$.
 
At $\mu\rho<0$ the situation is reversed, hence one can 
combine these two cases and cast the contribution of first two terms in 
(\ref{rkk-ne-0}) into the form \cite{Ab-St}
 
     \begin{mathletters}
     \label{rkk-ne-1}
     \begin{eqnarray}
     \chi_1(\br) &=&
     \frac {|\mu|}{ 8\pi t^2} \Phi_1(k_F^\ast r)
     \\ 
     \Phi_1(a) &=&
     J_0(a) Y_0(a) + J_1(a) Y_1(a)  
     \end{eqnarray}
     \end{mathletters}
with Bessel functions $J_n(x)$ and $Y_n(x)$
and 
the direction-dependent value of the effective Fermi momentum   
      \begin{equation}
       k_F^\ast = 
      \frac{\sqrt{|2\mu\rho|}}r  
      = \frac1a \sqrt{\left|\frac\mu t\cos(2\varphi) \right|} .
      \label{def-kF} 
      \end{equation}    
where we restored in the rhs
the lattice parameter $a$.   
Note that the expression (\ref{rkk-ne-1})
gives the  RKKY interaction 
for the cylindrical Fermi surface as well, in which case $k_F^\ast$ coincides 
with the conventional Fermi momentum \cite{rkky-anyD}.  The only
difference is in the general minus sign resulted from the sign-indefinite
property of the mass tensor, $det(\mass) = -1/(4t^2) < 0 $.

The asymptotes of this part of RKKY interaction 
under criteria (\ref{star2}) and (\ref{star3})
are as follows
     \begin{mathletters}
     \label{rkk-asymp1}
     \begin{eqnarray}
     \chi_1(\br) &=&
     \frac {|\mu|}{ 8\pi^2 t^2} \frac{\sin(2k_F^\ast r)}{(k_F^\ast r)^2}
     ,\quad k_F^\ast r \gg 1 
     \\ &=&
     \frac {|\mu|}{ 4\pi^2 t^2} \ln k_F^\ast r
     ,\quad k_F^\ast r \ll 1 
     \end{eqnarray}
     \end{mathletters}
Therefore in the limit of large distances the power-law 
decrease of $\chi_1(\br)$ is accompanied by oscillations with
the direction-dependent period $2k_F^\ast$, in accordance with usual 
expectations, while for the small $k_F^\ast r$ these oscillations 
are replaced by logarithmic singularity. 

The third term in (\ref{rkk-ne-0}) is the integral of the product of the 
Green's functions resulting from the 
regions of the FS with the different character of dispersion. 
As a result, one has the additional prefactor 
of the form $\exp(i {\bf Q}_0 \br) = (-1)^{x+y}$ (for integer $x,y$).  
The appearing  
``antiferromagnetic'' wave-vector $ {\bf Q}_0 = (\pi,\pi)$ merely 
connects two regions in the Brillouin zone, where the dispersion is close 
to the Fermi level.  
In this third term of Eq. (\ref{rkk-ne-0}) the discontinuity 
at $z = \mu$ exists for both signs of $\mu\rho$. The integration 
over $z$ can also be done \cite{Ab-St} and we obtain after some calculations :

     \begin{mathletters}
     \label{rkk-ne-2}
     \begin{eqnarray}
     \chi_2(\br) &=&
     e^{i {\bf Q}_0 \br}
      \frac {|\mu|}{ 4\pi^2 t^2} \Phi_2(k_F^\ast r)
     \\ 
     \Phi_2(a) &=&
      \frac{J_0(a) K_1(a) - J_1(a) K_0(a)}{a} 
     \end{eqnarray}
     \end{mathletters}

The asymptotes of this expression are 

     \begin{mathletters}
     \label{rkk-asymp2}
     \begin{eqnarray}
     \chi_2(\br) &=&
        e^{i {\bf Q}_0 \br}
     \frac { |\mu|\sqrt2}{ 4\pi^2 t^2} 
     \frac{\cos( k_F^\ast r)}{(k_F^\ast r)^2} e^{-k_F^\ast r}
     ,\quad k_F^\ast r \gg 1 
     \\ &=&
       e^{i {\bf Q}_0 \br}
     \frac {|\mu|}{ 4\pi^2 t^2} \frac 1{ (k_F^\ast r)^2 }
     ,\quad k_F^\ast r \ll 1 
     \end{eqnarray}
     \end{mathletters}

Let us discuss the significance of this second part of RKKY interaction. 
First, one sees from 
(\ref{rkk-asymp1}), (\ref{rkk-asymp2})
that in the far asymptotic region  $k_F^\ast r \gg 1$ the 
first term $ \chi_1(\br)$ dominates while $\chi_2(\br)$ is exponentially 
small. 
In particular, it explains why the  $\chi_2(\br)$ term
could not be obtained by the previous methods
\cite{rkky-ani} --- the RKKY interaction as expressed 
in Ref.\ \cite{rkky-ani} was a series in powers of $1/r$, and the exponential 
tail was evidently missing. 

On the contrary, the term $\chi_2(\br)$ starts to play a decisive  
role at smaller distances $k_F^\ast r \lesssim 1$, where it prevails
according to (\ref{rkk-asymp1}), (\ref{rkk-asymp2}).
We stress that the condition $k_F^\ast r > 1$ was essential for the 
previous theories, while our expressions 
are applicable at weaker
conditions (\ref{star2}),(\ref{star3}). Therefore the part $\chi_2(\br)$ 
represents the {\em intermediate asymptote} of the RKKY interaction in 
the nearly-nested situation. 
   
Second we note that $\chi_2(\br)$ has the antiferromagnetic sign-reversal 
character. This feature of the RKKY interaction at short distances
appears to be quite general. One can show for other types of dispersion
\cite{tobepub} that this kind of short range oscillations 
is always determined by the wave-vector connecting saddle points 
in the band structure.

To clarify these two items further we consider the case of the 
perfect nesting 
of the electronic spectrum, $\mu =0$, when the far asymptotic regime 
is not realized. We immediately see the disappearance 
of the first part of the RKKY interaction (\ref{rkk-ne-1}), while only 
this term could be obtained in the former methods of $1/r$ expansion.  
At the same time our second part of RKKY interaction (\ref{rkk-ne-2}) survives.
As a result we have  

     \begin{eqnarray}
     \label{perfect}
     \chi(\br) &=& 
       e^{i {\bf Q}_0 \br}
      \frac {1}{ 4\pi^2 t \,| x^2 - y^2| }
     \end{eqnarray}
This behavior indicates the log-squared singularity of the polarization 
operator on the antiferromagnetic wave vector ${\bf Q}_0$ discussed, 
e.g., by Dzyaloshinskii \cite{dzya}. 

To validate our analytical findings (\ref{rkk-ne-1}), (\ref{rkk-ne-2})
we performed the direct numerical 
calculation of the RKKY interaction 
on the square $|x|, |y| \leq 10$
with the spectrum given by (\ref{disp-nest}). 
The results for $\mu = 0.1$ and $t= 0.5$ are shown on the Figure 
\ref{fig:num}.
We plotted on the Figure \ref{fig:num}a 
the calculated value of the RKKY interaction versus the 
``distance'' in the saddle point metrics, $r^\ast = \sqrt{|x^2 - y^2|}$. 
According to this convention $k_F^\ast r = (k_F^\ast)_{max} r^\ast $, 
so we can show the results for the whole plane in a simplest manner; 
for the chosen parameters  $(k_F^\ast)_{max} = \sqrt{\mu/t} \simeq 0.45 $.
On the same plot we have drawn  the  curves 
$  \chi_1(k_F^\ast r) - \chi_2(k_F^\ast r)$ and 
$  \chi_1(k_F^\ast r) + \chi_2(k_F^\ast r)$,  
which are the predicted values of the interaction for the 
odd and even sites, respectively. No additional parameters were used. 

We see the remarkable agreement between the calculated points 
and the theoretical formulas.
As we expected at large distances the oscillations in the RKKY 
are observed while at smaller distances $r^\ast < \sqrt{t/\mu} \simeq 2.5$
the situation is changed. The interaction for the ``odd'' sites 
$x + y = 2n +1 $ (i.e.\ for $\br = (0,1), (1,2), (0,3)$\ldots) is of the 
antiferromagnetic (negative) sign.
The interaction for the ``even'' sites ($\br = (0,2), (1,3), (0,4)$ 
{\em etc.}) has a tendency to be ferromagnetic.  In both
cases the calculated points closely follow our curves
$ \chi_1(\br) \pm  \chi_2(\br)$ up to the interatomic distances. 
The RKKY interaction expected from the previously known 
expressions ( the term $  \chi_1(k_F^\ast r)$) is shown by a dashed 
line.  

To clearer represent the region of larger $r^\ast$ we multiplied the 
calculated points $\chi(\br)$ onto the appropriate values of $(r^\ast)^2$.  
The same was done for  the theoretical curves, the results 
are shown on the Figure \ref{fig:num}b.  We see again that 
at large  $r^\ast$ the interaction is characterized by 
the usually discussed asymptotic oscillations (\ref{rkk-asymp1}). 
At the same time the difference between the  
 ``odd'' and ``even'' sites is clear at the lower distances. 
  
Note that $\chi(\br)$ for the diagonal $x=\pm y$ is not present on the 
Fig.\ \ref{fig:num} and cannot be in principle compared to 
(\ref{rkk-ne-1}), (\ref{rkk-ne-2}) due to the criteria 
(\ref{star2}), (\ref{star3}); we discuss it also in the next Section.

\subsection{noninteger values of R}
%\label{sec:nonint}
 
Let us now extend our analysis for the case of non-integer 
values of $x, y$. One can easily note that now the factors of the 
type $\exp[{i\pi |x| sign(\omega )}]$ in (\ref{gf0pi}), (\ref{gfpi0}), 
 (\ref{appeq:Bzb2}) produce another source of 
discontinuity at $\omega = 0 $, in addition to the previously 
discussed one of the value $\sqrt{2\rho (\mu + i \omega)}$ . 
In particular both first and second terms in (\ref{rkk-ne-0}) 
acquire the factors $e^{ \pm 2i\pi |x|} \neq 1$ and 
$ e^{ \pm 2i\pi |y|} \neq 1$,
 respectively. 
Therefore both these terms now contribute although in a different 
manner.  

Consider first the case of $|x| > |y|$ and $\mu <0$, which means 
$\rho\mu <0$ and the closed character of the FS. 
A straightforward calculation \cite{Ab-St} shows then 
that the above expressions 
(\ref{rkk-ne-1}), (\ref{rkk-ne-2}) are generalized as follows :

     \begin{mathletters}
     \label{noninteger}
     \begin{eqnarray} 
     \chi_1(\br) &=&
      \frac {|\mu|}{ 8\pi t^2}      \left[
        \cos( 2\pi|x| )  \Phi_1(k_F^\ast r)
      \right. \nonumber \\  &&  
     +   \sin( 2\pi|x| )   \Phi_3^{(1)}(k_F^\ast r)
      \\  && \left.
     + \sin( 2\pi|y| ) \Phi_3^{(2)}(k_F^\ast r) 
     \right] \nonumber \\ 
     \chi_2(\br) &=&
      \frac {|\mu|}{ 4\pi^2 t^2}      \left[
        \cos( \pi|x| + \pi|y|)  \Phi_2(k_F^\ast r)
      \right. \nonumber \\  && \left.  
     +   \sin( \pi|x| + \pi|y|)     \Phi_4(k_F^\ast r)
     \right] 
     \end{eqnarray}
     \end{mathletters}
with the functions

     \begin{mathletters}
     \label{nonint-fu}
     \begin{eqnarray} 
     \Phi_3^{(1)}(a) &=&
      \frac12 [ Y_0^2(a) - J_0^2(a) + Y_1^2(a) - J_1^2(a) ]
      \\
     \Phi_3^{(2)}(a) &=&
      \frac2{\pi^2} [  K_0^2(a) - K_1^2(a)  ]  
      \\
     \Phi_4(a) &=&
      \frac{Y_0(a) K_1(a) - Y_1(a) K_0(a)}{a} 
     \end{eqnarray}
     \end{mathletters}
The different terms appeared in (\ref{noninteger}) have different 
significance at large and small $k_F^\ast r$. 

In the far asymptotic regime  $k_F^\ast r \gtrsim 1$ 
the terms $\Phi_3^{(2)}, \Phi_2, \Phi_4$ are  exponentially small and 
we find :

     \begin{eqnarray} 
      \chi(\br) \propto  
      \frac{\sin (2\pi|x| - 2k_F^\ast r) }
      { (k_F^\ast r)^2} , \quad k_F^\ast r \gtrsim 1
      \label{far}
     \end{eqnarray}
The period of oscillation in the above expression  
corresponds to the notion of the calipering points 
on the FS. \cite{rkky-ani} 
We remind that these are the points where the direction 
of normal to the 
Fermi surface is (anti)parallel to the direction of $\br$.
In other words, 
the normal to the FS coincides with the direction of the Fermi velocity 
${\bf v}  = (v_x, v_y)$ and it is parallel to $\br = r (
\cos \varphi, \sin \varphi)$  provided  
      \begin{equation} 
      \label{calpoint}
      v_x \sin \varphi = v_y \cos \varphi.
      \end{equation} 
Near the saddle points $(\pm\pi,0)$ one has  $v_x / v_y = - k_x / k_y $ and 
$ k_x^2 - k_y^2 = -\mu/t $. Therefore the calipering points, satisfying 
the condition (\ref{calpoint}), are given by $\widetilde \bk^c = 
\pm ( \cos \varphi, - \sin \varphi) [-(t/\mu) \cos 2 \varphi ]^{-1/2}$
near the points $(\mp\pi,0)$, respectively. 
 We measured 
the wave-vectors from the saddle points, therefore the true 
caliper of the fermi surface is given by the vector 
$\bk ^c = (2\pi, 0 )  - 2 \widetilde \bk^c$. The scalar product $\bk^c \br$  
is exactly what one finds in  
the Eq.(\ref{far}) since    $\widetilde \bk^c \br = 
 r \sqrt{-\mu \cos(2\varphi) /t} \equiv  k_F^\ast r$.

At the smaller distances $k_F^\ast r \leq 1$ these are the terms 
$\Phi_3^{(1)}, \Phi_3^{(2)},\Phi_2 $ , which determine the main contribution 
to  $\chi(\br)$. In this case one obtains : 

     \begin{eqnarray} 
      \chi(\br) &\simeq& 
      \frac{\pi \cos (\pi|x| +\pi|y|) 
       + \sin 2\pi|x|  - \sin 2\pi|y| }
      { 4 \pi^3 t |x^2 - y^2 | } ,
      \label{near}\\              
        && \quad k_F^\ast r \lesssim 1
      \nonumber
     \end{eqnarray}
We see again that the interaction has a commensurate period of 
oscillations, although the oscillations for non-integer $\br$ 
are not described by a unique factor as it was in Eqs.(\ref{rkk-asymp2}b),
(\ref{far}).

\section{the flat parts of the spectrum}
\label{app:flat}

Let us discuss here the contribution to the RKKY interaction produced 
by the flat parts of the spectrum (\ref{exp4}). First we observe that 
the formalism developed in the main part of the paper cannot be applied 
to the vicinities of the points $(\pm\pi/2,\pm\pi/2)$ since all the 
components of the mass tensor are infinite at these points. This 
very special case should be treated separately; one can also distinguish 
here the regions of intermediate and far asymptotes. 
At the intermediate distances $r \lesssim |t/\mu|$ we obtain the 
  Green's function 
from the vicinities of  $( \pi/2, \pi/2)$ and  $( -\pi/2, -\pi/2)$ in 
the form

     \begin{equation}
     G_{ (\pi/2,\pi/2)}(i\omega, \br ) =
     \frac{ e^{-i\alpha} }{ 2 t} 
     \delta_\kappa( x- y)   
     \exp{ \left[ i |x|(\pi + z/2t ) sign(\omega ) \right] }     
     \label{gf-cBz}
     \end{equation}
and the corresponding Green's function from the 
vicinities of  $( \pi/2, -\pi/2)$ and  $( -\pi/2, \pi/2)$ is obtained 
from this expression by the replacement  $y \to - y$.
 The function $\delta_\kappa(x)$ in  (\ref{gf-cBz}) 
 has the $\delta-$function-like properties and  is defined by  
      \begin{equation} 
     \delta_\kappa(x) =
     \frac{\sin \kappa x}{\pi x}, \quad \kappa \sim \frac1a
     \label{appeq:delta}
     \end{equation}
  We wish to point out that the power-law
decrease of $\delta_\kappa(x)$ at large $x$ stems from our assumption that 
the absence of dispersion along $x$ is lost abruptly at $|k_x| > \kappa$. 
In fact the expression  (\ref{appeq:delta}) is the Fourier transform of
$\theta(\kappa - |k_x|)$. In general the dependence of dispersion on $k_x$ is 
much smoother ; as a result, the decay of $\delta_\kappa(x)$ at large $x$
should be much faster, while  the $\delta-$function-like 
property preserves.

We see that the above Green's function has a sizeable values only in a band 
$|x - y| \lesssim 1$. Outside this domain   the principal 
contribution to the total $G(i\omega, \br)$  (\ref{G-total})  
is delivered by $G_{(0,\pi)}(i\omega, \br )$ and  $G_{(\pi,0)}(i\omega, 
\br )$, Eqs.\ (\ref{gf0pi}) and (\ref{gfpi0}).
Particularly it means (see Fig.\ref{fig:regions}) that the terms 
of the type $ G_{ (\pi/2,\pi/2)} G_{(0,\pi)} $ in the expression 
(\ref{rkk-ne-0}) should not be considered.  

As a result the contributions to the RKKY interaction from the flat parts 
of FS  acquire the following form :  

     \begin{eqnarray}
     \chi_{flat} (\br ) &=&
     \frac{\cos x(2\pi + \mu/t )}{ 4\pi t |x|} 
     \left[ \delta_\kappa^2( x- y)  + \delta_\kappa^2( x+ y)  
       \right],
     \label{chi-flat} \\
      && \quad |x|,|y| \lesssim |t/\mu|
     \nonumber
     \end{eqnarray}
here two terms in the square brackets correspond to the different 
regions in the $\br$-space.
We see that this part of interaction which is present along the diagonals
is slowly decaying as $1/r$. The amplitude of it, according to 
(\ref{appeq:delta}) has the model cutoff parameter $\kappa^2$. Hence we 
cannot directly compare this part of RKKY  with the results of our
numerical calculations, although the overall $1/r$ dependence of RKKY 
interaction along the diagonals and slow oscillations are verified by 
the numerical data as well. At small integer values of 
$x=y$ the RKKY term  (\ref{chi-flat}) corresponds to the 
ferromagnetic sign of the interaction between the localized moments. This 
behavior however does not define the particular type of magnetic ordering     
and it is the term (\ref{perfect}) which determines it. 

At extremely large distances close to diagonal $r\gg |t/\mu|$ the dropped 
$k^3-$terms in the expansions (\ref{exp3}), (\ref{exp4}) 
become important. Hence the spectrum becomes essentially two-dimensional, 
with the corresponding
change in the character of RKKY. Near the diagonals 
$\varphi = \pm\pi/4 + \phi $ one has : 

     \begin{equation}
     \chi_{flat} (\br ) =
     -\frac{\sin |x|(2\pi + \mu/t (1- \phi^2/\phi_0^2) )}
      { 4\pi^2 |\mu| x^2  [1+ \phi^2/\phi_0^2 ]  }      
           ,\quad |x| \gg \frac t {|\mu|}
     \label{chi-flat-far}
     \end{equation}
 This far asymptote of RKKY interaction from the flat 
parts of dispersion holds in the narrow sectors near the diagonals
$  |\phi| \lesssim \phi_0 = \sqrt2 |\mu/4t|$,  
( cf. (\ref{star2}), (\ref{star3}) and Fig.\ \ref{fig:regions} ). It has 
 the $1/r^2$ dependence while its period of oscillations correponds
to the notion of calipering points discussed above.

\section{concluding remarks}

It is worthwhile to compare our expressions for the Green's 
function with previous results. 
It was observed  (see e.g. \cite{economou}) that for 
the tight-binding spectrum (\ref{disp-nest})  
one can use the recurrence relations to express the 
value of $G(\omega, \br)$ in terms of $G(\omega, 0)$.       
It was noted also however that using these relations   
one meets the numerical 
instabilities at large $r$. Alternatively  
$G(\omega, \br)$  can  be estimated
at large $r$ by the steepest descent method. The solution 
obtained by this latter method corresponds to the asymptotes of Eqs. 
(\ref{g-mcdo}), (\ref{K2H}) for the 
case of large $\sqrt{2\mu\rho} = k_F^\ast r$. 
In this sense our expressions extend the 
previous findings for the Green's function and provide the analytical
formulas in the region of the intermediate 
distances $1\lesssim r \lesssim 1/k_F^\ast$.  

Let us briefly discuss here the role 
  of finite temperatures for our treatment. 
In this case instead of the integral (\ref{rkk-ne-0})
one considers the sum over the Matsubara frequencies 
(\ref{rkk-inter}) with the Green's functions given by (\ref{gf0pi}),
 (\ref{gfpi0}). 
With the use of analytical continuation, this sum can be represented 
as the integral over the real axis of $\omega$. 
One can note however directly from the 
form of the Green's functions that the
effect of finite temperatures
is important when $T$ exceeds the effective 
chemical potential $\mu$. At large distances $ r \gtrsim \xi = 
(k_F^\ast)^{-1}\sqrt{\mu/T}\propto \sqrt{t/T}$ we have 
the RKKY interaction exponentially suppressed. \cite{fnote3}
The opposite case $ r \lesssim \xi$
corresponds essentially to the case $\mu = 0 $ described by the 
Eq.(\ref{perfect}). Therefore the far asymptote of RKKY leading to 
possible incommensurate magnetic ordering is absent in this case and we 
remain with the only tendency to commensurate AF order.  

In conclusion we found the closed analytic expressions for the  
RKKY interaction in a layered metal with nearly nested   
Fermi surface. Along with the usual $2k_F$-oscillations 
realized at far distances we demonstrate the existence 
of the intermediate asymptote of the interaction. This latter
asymptote has the commensurate AF type of oscillations and is the only 
term surviving at the exact nesting. 
We show that our analytical formulas are in the good accordance 
with the numerically found values of interaction in a range 
up to near interatomic distances.

%%%%%%%%%%%%%%%%%%%%%%%%%%%%%%%%%%%%%%%%%%%%%%%%%%%%%%%%%%%%%%%%%5
%%%%%%%%%%%%%%%%%%%%%%%%%%%%%%%%%%%%%%%%%%%%%%%%%%%%%%%%%%%%%%%%%5
%%%%%%%%%%%%%%%%%%%%%%%%%%%%%%%%%%%%%%%%%%%%%%%%%%%%%%%%%%%%%%%%%5
%%%%%%%%%%%%%%%%%%%%%%%%%%%%%%%%%%%%%%%%%%%%%%%%%%%%%%%%%%%%%%%%%5

\acknowledgements

We thank A. Furrer, M. Kiselev, F. Onufrieva, P. Pfeuty 
for useful discussions.  
The financial support of this work from the Russian
Foundation for Basic Researches (  Grant No.\
96-02-18037-a ) and Russian State Programme for Statistical Physics 
is acknowledged. 
One of us (D.N.A.) thanks  LNS at ETH\&PSI for the hospitality extended 
to him during his visit there.

\appendix

\section{the stationary point at the zone boundary}
\label{app:bzb}

Let us consider the case of the stationary point lying on the
boundary of the Brillouin zone. We discuss it on the example of
the sum of the points $  (\pi,0)$ and $(-\pi,0)$ 
for the tight-binding form of
dispersion (\ref{disp-nest}).  The integration
over $k_y$ meets no difficulties, while the domains of
integration over $k_x$, namely $(-\pi, -\pi+\kappa)$  and
$(\pi-\kappa, \pi)$, could be combined as follows :

     \begin{eqnarray}
     G_{ZB} (i\omega, x) &=&
     e^{-i\pi x}
     \int_0^\kappa \!
     \frac{ dk_x }{2\pi} \!
     \exp \!\left[
     i k_x x - \tau e^{i\alpha} \frac{ k_x^2 }{2m_x}
     \right]
     \nonumber \\ &&
     + \left( x \to -x \right)
     \nonumber \\
     &=&  \left.
     A \left[  \cos(\pi x)  {\rm Erf}\left(
     \frac{k \gamma}2 - i\frac{x}\gamma
     \right)
     \right|_{k=-\kappa}^\kappa \right.
     \nonumber \\ &&
     \left. + 2i   \sin (\pi x) \,
      {\rm Erf}\left( - i\frac{x}\gamma
     \right) \right]
     \label{appeq:Bzb}
     \end{eqnarray}
with
 
      \begin{eqnarray}
     \label{appeq:A}
     A &=& \frac1{\pi\gamma}
     \exp\left[ - e^{-i\alpha} \frac{m_xx^2}{2\tau}
     \right]    ,\\ 
      \gamma &=&  \sqrt{e^{i\alpha}\frac{2\tau}{ m_x}}, \quad
      Re(\gamma) > 0  .
     \nonumber 
     \end{eqnarray}
The demand for the argument $\pm \kappa \gamma/ 2 - i{x}/\gamma$ in 
  (\ref{appeq:Bzb}) to be large corresponds to the criteria of  
applicability of our expressions   (\ref{star2}), (\ref{star3}) 
and we do not discuss it here. 
Similar to the above treatment at $ k_F^\ast r \gg 1 $ one has 
$\tau \simeq  k_F^\ast r/(2\mu)$ and  $|x/\gamma| \geq  k_F^\ast r \gg1$. 
At  $k_F^\ast r \lesssim 1$ we saw that 
$|\rho|\lesssim \tau \lesssim |z|$ ; on the other hand the values of 
$|z|$ at which the integral (\ref{rkk-ne-0}) saturated 
were of order of $1/|\rho|$. Therefore we can let  $\tau\sim |\rho| $
and obtain $|x/\gamma| \sim \sqrt{|x^2m_x /\rho|}\geq1$. Hence in both 
cases the argument of the second error function in  (\ref{appeq:Bzb}) is 
large enough, while its sign coincides with the sign of a product 
$- x\omega m_x$. Therefore the Green's function has a 
following form : 

     \begin{eqnarray}
     G_{zb}(i\omega, x) =
     \sqrt\pi  A \exp[ {-i\pi |x| sign(\omega m_x) } ]
     \label{appeq:Bzb2}
     \end{eqnarray}
This result means that the Gaussian value of integral is
attained in one of the above domains of integration. Note that
$G_{zb}(i\omega, x) $ is an even function of $x$ ; it is a
manifestation of the reflecting property of the Brillouin zone
boundary \cite{aaa}. For a particular choice of our spectrum 
(\ref{exp2}) we have $m_x = -1 /2t<0$ and  thus the formula 
(\ref{gfpi0}).

% now the references. delete or change fake bibitem. delete next three
%   lines and directly read in your .bbl file if you use bibtex.

%
%  Figure 1 
% 
\begin{figure}
\centerline{\epsfxsize=8cm \epsfbox{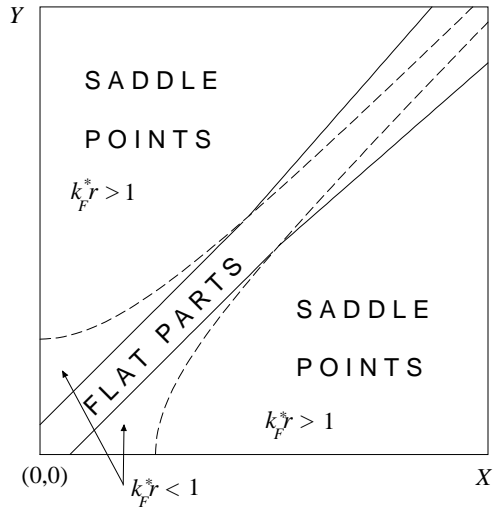}} 
\caption{ The regions of applicability of the different 
Green's functions in the sector $x>0$, $y> 0$ of the $\br$-space. 
The contributions from the 
saddle points of the spectrum  are given by 
  (\ref{gf0pi}), (\ref{gfpi0}) and are defined away 
from the diagonal. The expression (\ref{gf-cBz}) 
provides the contribution from the flat parts of dispersion. 
The solid line is the border between the applicability regions of 
corresponding equations. At the same plot the curve $k_F^\ast r =1 $
is shown by the dashed line.  
 \label{fig:regions}  }
\end{figure}

%
%  Figure 2 
%
\begin{figure}
\centerline{\epsfxsize=8cm \epsfbox{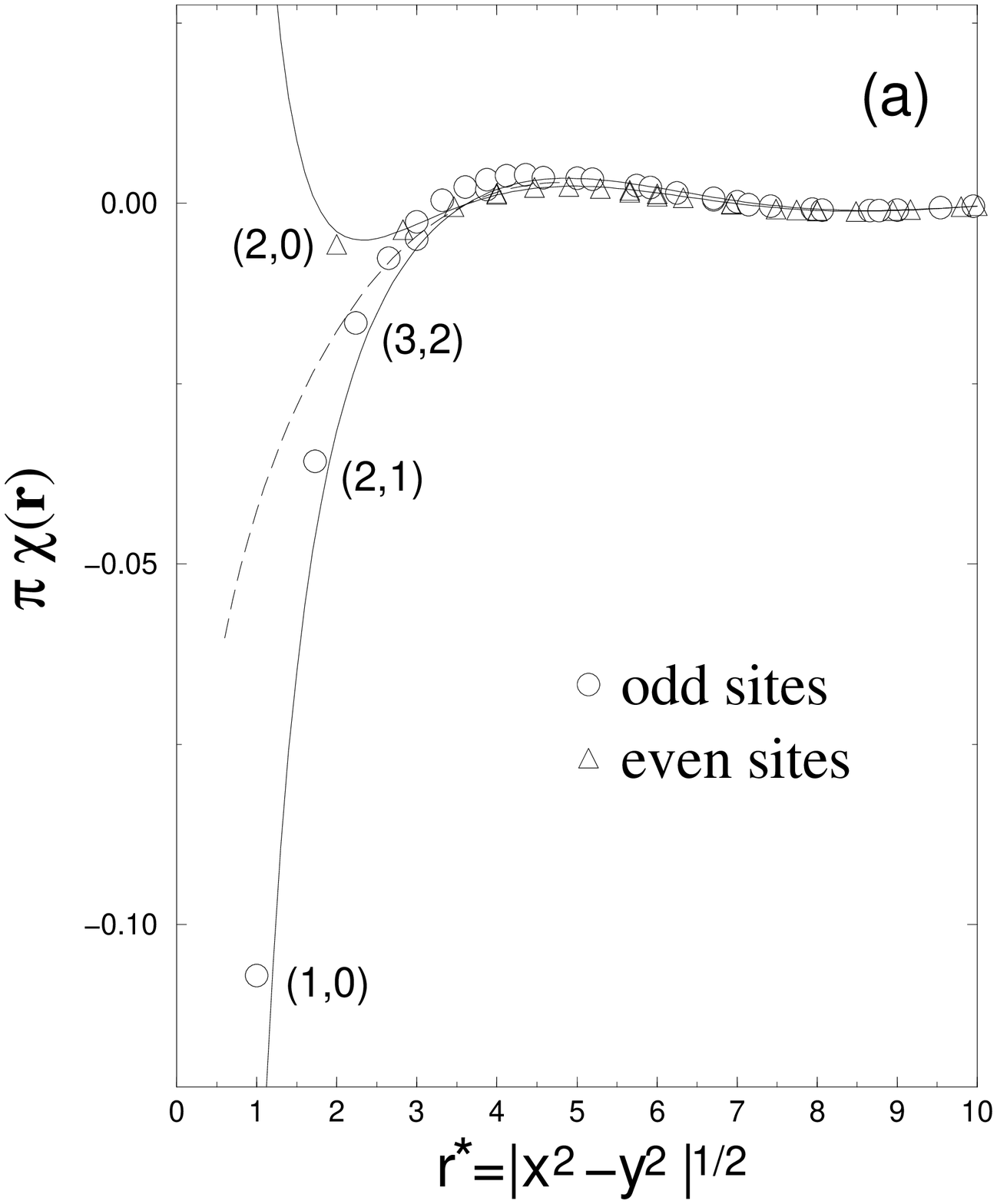}} 
\centerline{\epsfxsize=8cm \epsfbox{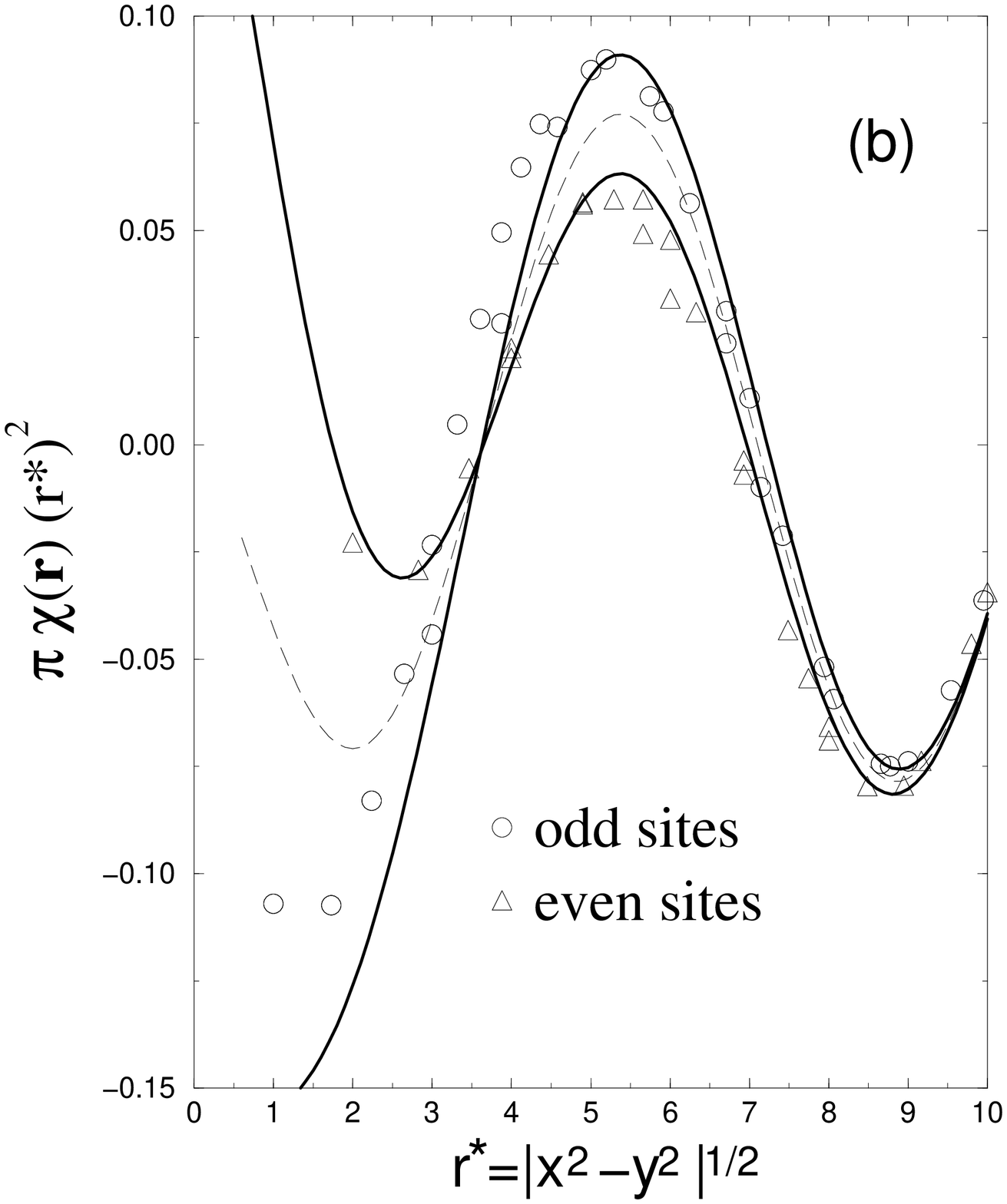}}
\caption{ a) The calculated RKKY interaction for the tight-binding 
spectrum (\ref{disp-nest}) with $\mu = 0.1, t=0.5$
is shown along with the theoretically obtained curves. 
b) The same quantity plotted in a modified way, for clearer 
 representation of the region of larger $r^\ast$. 
The actual coordinates of some sites are indicated. 
One can note the remarkable accordance between the numerical findings 
and the analytical results in the intermediate region, where the
curves for ``odd'' and ``even'' sites differ visibly.    
The asymptotic solution for RKKY extended to shorter distances 
is shown by a dashed line for comparison.
 \label{fig:num}  }
\end{figure}


\begin{references}
\bibitem{fnote2} Note that one can neglect the $\omega-$dependence of $\tau$
here since the subsequent integration over $\omega$ in (\ref{rkk-ne-0})
 saturates over  $\omega \lesssim \mu/(k_F^\ast r) \ll \mu$.
\bibitem{fnote3} Note that for spherical FS 
one would have $T \ll \mu $ and the temperature correlation 
length $ \xi \propto \mu/T$
\end{references}
\end{document}